\documentclass[aps,prl,twocolumn,groupedaddress,nofootinbib,showpacs]{revtex4-1}

\usepackage{graphicx}
\usepackage{mathptmx}      % use Times fonts if available on your TeX system
\usepackage{subfigure}
\usepackage{color}

\begin{document}

\title{Equilibration and Thermalization of Classical Systems}

\author{Fengping Jin}
\affiliation{Institute for Advanced Simulation, J\"ulich Supercomputing Centre,\\
Forschungzentrum J\"ulich, D-52425 J\"ulich, Germany}
\author{Thomas Neuhaus}
\affiliation{Institute for Advanced Simulation, J\"ulich Supercomputing Centre,\\
Forschungzentrum J\"ulich, D-52425 J\"ulich, Germany}
\author{Kristel Michielsen}
\affiliation{Institute for Advanced Simulation, J\"ulich Supercomputing Centre,\\
Forschungzentrum J\"ulich, D-52425 J\"ulich, Germany}
\affiliation{RWTH Aachen University, D-52056 Aachen, Germany}
\author{Seiji Miyashita}
\affiliation{Department of Physics, Graduate School of Science,\\
University of Tokyo, Bunkyo-ku, Tokyo 113-0033, Japan}
\affiliation{CREST, JST, 4-1-8 Honcho Kawaguchi, Saitama, 332-0012, Japan}
\author{Mark Novotny}
\affiliation{Department of Physics and Astronomy, Mississippi State University, Mississippi State, MS 39762-5167, USA}
\affiliation{HPC$^2$ Center for Computational Sciences, Mississippi State University, Mississippi State, MS 39762-9627, USA}
\author{Mikhail I. Katsnelson}
\affiliation{Institute of Molecules and Materials, Radboud University of Nijmegen,
NL-6525ED Nijmegen, The Netherlands}
\author{Hans De Raedt}
\affiliation{Department of Applied Physics, Zernike Institute for Advanced Materials,\\
University of Groningen, Nijenborgh 4, NL-9747AG Groningen, The Netherlands}

\date{\today}

\begin{abstract}
It is demonstrated that the
canonical distribution for a subsystem of a closed system
follows directly from the solution of the time-reversible Newtonian equation of motion
in which the total energy is strictly conserved.
It is shown that this conclusion holds for both integrable or nonintegrable systems
even though the whole system may contain as little as a few thousand particles.
In other words, we demonstrate that the canonical distribution holds
for subsystems of experimentally relevant sizes and observation times.
\end{abstract}

\maketitle

Boltzmann's postulate of equal \textit{a-priori} probability is the cornerstone of statistical mechanics.
This postulate eliminates the difficulties of deriving the statistical properties
of a many-body system from its dynamical evolution by
introducing a probabilistic description in terms of the (micro)canonical ensemble.
In classical mechanics, the equivalence of these two fundamentally
different levels of description remains elusive~\cite{KUBO85,GREI97}.

Although considerable progress has been made through the development of ergodic theory,
more than hundred years after its conception
a direct demonstration that the equal \textit{a-priori} principle
or, equivalently, the canonical distribution follows directly from the dynamical behavior
of a finite number of particles is still missing~\cite{KUBO85,GREI97}.

The discovery of ``deterministic chaos'' has made a huge impact on old discussions of the relations
between classical and statistical mechanics~\cite{SCHU95,ZASL85,SAGD88,PRIG80}.
The irreversibility in the macroworld is often related with unstable motion in generic mechanical systems
characterized by positive Lyapunov exponents and a non-zero Kolmogorov entropy.
Integrable systems are considered from this point of view as exceptional (not to say pathological)
and irrelevant for the problem of justification of statistical physics.
The famous Fermi-Pasta-Ulam paradox~\cite{FERM55,FORD92,GALL08,PORT09} and its interpretation in terms of closeness of their
model to the completely integrable KdV model (see Ref.~\cite{ZASL85}) has emphasized (probably, overemphasized) this
point. Indeed, there is no tendency to equilibration in a system of noninteracting entities (particles
of an ideal gas, normal modes in harmonic oscillator systems, solitons in KdV systems, etc.). However, if we
discuss an isolated Hamiltonian system there is no way to obtain the statistical mechanical behaviour (in particular
irreversibility), not even for a system with chaotic motion.
An alternative is to consider an {\it open} system.
For the open system, the role of integrability should be
discussed in a different context. The integrability of the isolated Hamiltonian system implies that its Liouville
operator $L$ is diagonal in the representation of angle-action variables~\cite{SAGD88}. If we choose a subsystem $S$
of the isolated integrable system which is also integrable, then its Liouville operator $L_S$ is diagonal
in the angle-action variables of the subsystem. However, these variables can be different
from those of the isolated Hamiltonian system.
$L$ and $L_S$ do not commute and cannot be diagonalized {\it simultaneously}. It is not clear \textit{a-priori}
whether in this situation the subsystem $S$ can equilibrate or not, and what the conditions for this
equilibration are. Surprisingly, this very natural issue is not clarified yet and thus requires additional
studies.

Here we present clear and unambiguous evidence that the
canonical distribution for $S$
follows directly from the solution of the time-reversible Newtonian equation of motion.
Furthermore, it is shown that the energy of the subsystem and of the environment equilibrate
on a relatively short, microscopic time scale, even if the whole system
contains only of the order of a thousand degrees of freedom.
The key feature of our demonstration is that we follow the time evolution
of a closed (isolated) system with a fixed energy and consider only a subsystem $S$ of the closed system.
The time evolution of the entire system is obtained by solving the Newtonian equations of motion
for typical initial states.
We do not perform ensemble averaging nor do we invoke
arguments based on (non)ergodicity~\cite{HEMM58,MAZU60,FORD63,KUBO85,GREI97}
or on the thermodynamic limit~\cite{KUBO85,GREI97}.
Nevertheless, we observe that $S$ is governed by the canonical distribution.

Our demonstration is not a mathematical proof but is based on the exact numerical solution
of what is perhaps the simplest of all interacting many-body systems: a one-dimensional harmonic oscillator model of a solid.
By solving the Newtonian equation of motion of the whole, integrable system and analyzing
only a single trajectory of the subsystem in phase space,
we show that the number of times that the subsystem is observed to posses a certain energy
is distributed according to canonical ensemble theory,
implying for example that within the subsystem equipartition holds~\cite{GIBB02,TOLM27,ULIN08,EAST10}.
Repeating the analysis for a one-dimensional model
of classical magnetic moments which is known to exhibit chaotic behavior~\cite{STEI09},
it is found that this conclusion does not depend on whether or not the motion is chaotic.
For both models, the distributions extracted from the single-trajectory Newtonian dynamics are
found to be in excellent agreement with the corresponding
results of microcanonical Monte Carlo simulations.

\begin{figure}
\includegraphics[width=8cm]{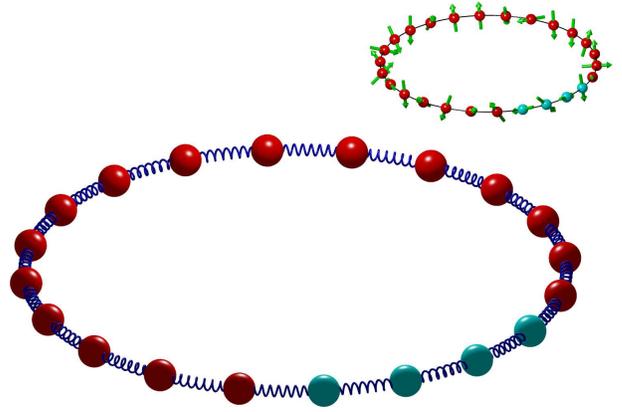}
\caption{
Picture of the harmonic oscillator (large) and magnetic moment (small) models,
subject to periodic boundary conditions.
Particles are connected by harmonic springs or carry a magnetic moment
that interacts with its nearest neighbors.
The blue (red) colored particles belong to the subsystem (environment).
}
\label{model}
\end{figure}

As a first example, we consider the most basic classical model for the vibration in a solid,
namely a set of particles of mass $m$,
arranged on a ring and connected to their two neighbors by harmonic springs, see Fig.~\ref{model}.
The Hamiltonian of the system reads
\begin{equation}
\label{hamiltonian}
H=\sum_{i=1}^N \frac{p_i^2}{2m} + \frac{m\Omega^2}{2}\sum_{i=1}^N (x_i-x_{i+1})^2,
\end{equation}
where $x_i$ and $p_i$ are the displacement and momentum of the $i$th oscillator,
$m\Omega^2$ is the spring constant and $N$ is the total number of particles.
For convenience, $m$ and $\Omega$ are set to $1$ and all quantities
such as momenta and displacements are taken to be dimensionless.
The positions are constrained to lie on a circle.
Changing to normal-mode coordinates $\{P_k,X_k\}$,
the Hamiltonian reads $H=(1/2)\sum_{k=0}^{N-1} \left( P_k^2 + \omega_k^2 X_k^2 \right)$
where $\omega_k=2|\sin \pi k/N|$.
The motion of each set of coordinates $\{P_k,X_k\}$ is described by a single sinusoidal oscillation
and is decoupled from the motion of all other sets.
Clearly, this classical system is integrable which allows us to compute numerically
the coordinates and momenta of the oscillators without introducing systematic or cumulative errors.
To this end, we transform the set of values of $\{x_1,\ldots,x_N,p_1,\ldots,p_N\}$ at time $t=0$
to their corresponding normal-mode values, use the simple sinusoidal dependence of the latter
to obtain their values at any point $t$ in time, and use the inverse transformation
to find the values of $\{x_1,\ldots,x_N,p_1,\ldots,p_N\}$ at time $t$.
This whole procedure is numerically exact, up to machine precision.

As a second example, we consider a ring of classical magnetic moments,
their total energy given by the Hamiltonian~\cite{FISH63}
\begin{equation}
H=-J\sum_{i=1}^N \mathbf{S}_i\cdot\mathbf{S}_{i+1},
\label{hamiltonian2}
\end{equation}
where $\mathbf{S}_i$ is a $3$-dimensional unit vector, representing the magnetic moment of a particle at lattice
site $i$, $J$ defines the energy scale which we set equal to $1$ in our numerical work and $N$ is the total number of moments.
The equation of motion of these moments reads
\begin{equation}
\label{eom2}
\frac{d}{dt}\mathbf{S}_i=\frac{\partial H}{\partial \mathbf{S}_i}\times\mathbf{S}_i
=-J\mathbf{S}_i\times\left(\mathbf{S}_{i-1}+\mathbf{S}_{i+1}\right)
,
\end{equation}
which obviously is nonlinear.
Nevertheless, Eq.~(\ref{eom2}) admits a harmonic-wave solution
$\mathbf{S} _i (t) = (\mathbf{a}\cos \theta_i + \mathbf{b} \sin \theta_i)\cos \phi + \mathbf{c}\sin \phi$,
where $\theta_i=ip-\omega t$, $\omega = 2(1-\cos p)\sin\phi$, $\phi$ and $p$ are real constants,
and $(\mathbf{a},\mathbf{b},\mathbf{c})$ form a right-handed set of orthogonal unit vectors~\cite{LAKS76,ROBE88}.
More generally, Eq.~(\ref{eom2}) has simple analytical solutions for $N=2$ and $N=3$~\cite{STEI09}.
The motion of $N=4$ magnetic moments arranged on a ring is regular~\cite{STEI09}.
For $N>4$, the system exhibits chaotic motion~\cite{STEI09},
except for special initial conditions such as the spin-wave and soliton solutions~\cite{SCHM11}.
We integrate the nonlinear equations of motion using a
fourth-order Suzuki-Trotter product-formula method
which conserves (1)
the volume of the phase space, (2) the length of each magnetic moment
and (3) the total energy~\cite{KREC98}.
Due to the chaotic character of the motion of the magnetic moments,
their trajectories are unstable with respect to
rounding and time-integration errors.
Nevertheless, the numerical method used guarantees that the motion
of the magnetic moments strictly conserves the energy,
as required for a microcanonical ensemble simulation.

\begin{figure}
\includegraphics[width=9cm]{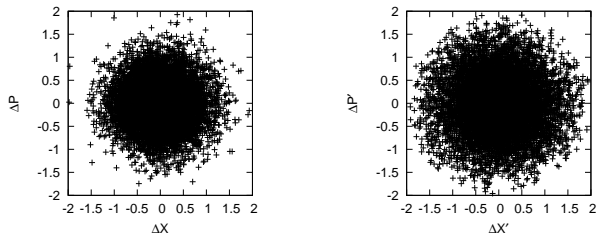}
\caption{
Left: Poincar\'e-type map of the differences
$\Delta X=(x_{0}-x_{1})/2$ and $\Delta P= (p_{0}-p_{1})/2\sqrt{2}$
for a ring of $N=512$ oscillators.
This system is integrable.
Right: Poincar\'e-type map of the differences
$\Delta X'=S^x_1-S^x_2$ and $\Delta P'=S^z_1-S^z_2$
for a ring of $N=512$ magnetic moments.
This system is chaotic~\cite{STEI09}.
Each graph contains 5000 points.
}
\label{map}
\end{figure}

In our numerical work, the initial values of the
coordinates or magnetic moments are
generated by standard Monte Carlo methods~\cite{LAND00},
see the Supplementary Information.

In the normal-mode representation, each of the oscillators traces out
an ellipse in the $(X_k,P_k)$ plane.
However, in the original coordinates, this simplicity is lost
as illustrated by the Poincar\'e map shown in Fig.~\ref{map}(left).
Clearly, it is difficult to detect some regularity in this map.
Moreover, this map is very similar to Fig.~\ref{map}(right), the Poincar\'e-type map of a
one-dimensional classical model of magnetic moments.
This similarity exists in spite of the fact that
the oscillator system is not chaotic.

The whole system is divided into two parts, a subsystem with $N_S$ particles
and an environment with $N_E=N-N_S$ particles.
The Hamiltonian is written as $H=H_S+H_E+H_{SE}$, where $H_S$ ($H_E$)
denotes the energy of the subsystem (environment) and $H_{SE}$ denotes the energy due to the
interaction of the subsystem with the environment.
Thus, in the case of particles arranged on a ring,
$H_S$ and $H_E$ are open chains of particles,
and $H_{SE}$ only contains two terms of the form $(x_i-x_{i+1})^2$
or $\mathbf{S}_i\cdot\mathbf{S}_{i+1}$ for the oscillator and magnetic system, respectively.

\begin{figure}
\includegraphics[width=8cm]{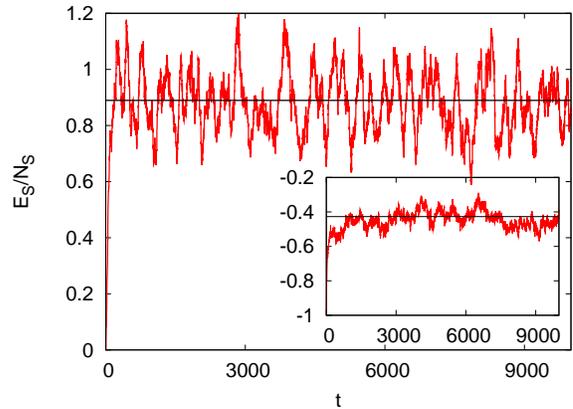}
\caption{
Time evolution of the energy per particle $E_S/N_S$, as obtained from a chain of $N_S=80$ particles
embedded in a ring of $N=512$ particles.
Initially, the subsystem of $N_S=80$ particles is put in its ground state and
the configuration of the environment of $N_E=N-N_S=432$ particles is
drawn randomly from its canonical distribution at temperature $T=1$.
Red line: energy per particle of the subsystem;
horizontal black line: energy per particle of the whole system.
Main figure: the energy of the oscillators in the subsystem quickly
relaxes to the average energy of the whole system.
The inset shows the corresponding data for the spin system.
}
\label{equilibration}
\end{figure}

We first study the dynamic evolution of the subsystem when it is brought in contact with the environment.
Initially, using one of the procedures described in the Supplementary Information,
the subsystem and environment are prepared such that they have a different energy.
As the whole system evolves in time, strictly conserving the total energy,
we monitor the energy of the subsystem as a function of time.
Some representative results are shown in Fig.~\ref{equilibration},
for both the harmonic oscillator and magnetic moment model.
From Fig.~\ref{equilibration}, it is clear that for both models, the
energy of the subsystem $E_s$
rapidly approaches the average energy of the whole system.

\begin{figure}
\includegraphics[width=8cm]{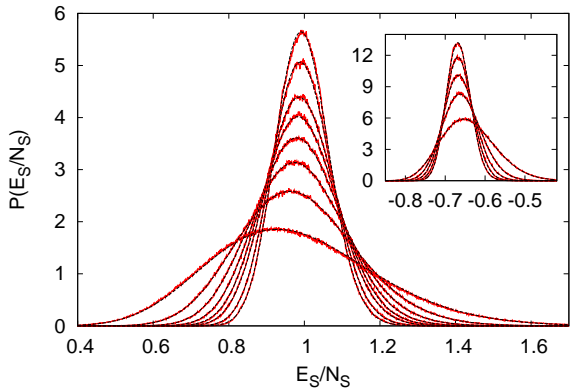}
\caption{
The distribution $P(E_S)$ as a function of the energy per particle $E_S/N_S$
for different sizes $N_S$ of the subsystem, as obtained from
the solution of Newtonian equations of motion for the oscillator system.
Red lines: $N_S=20, 40, 60, 80, 100, 120, 160, 200$ (broad to narrow) and $N=65536$.
Black lines: probability distribution predicted by statistical mechanics
Eq.~(\ref{cd2}) using all terms up to $(E-E^\ast)^8$.
In all cases, the initial energy of the environment
corresponds to a temperature $T=1$ and the number of samples is $4\times10^6$.
The inset shows the corresponding data for the spin system
for $N_S=20, 40, 60, 80, 100$ and $N=2000$, with the initial energy
of the environment corresponding to $T=0.329$
and the number of samples is $10^6$.
}
\label{distribution}
\end{figure}

Having established that the Newtonian equation of motion
drives the subsystem and environment to a common equilibrium state,
the next step is to study the distribution of the subsystem energy.
After the relaxation to the equilibrium state, we monitor the energy $E_S$ of the subsystem
at regular time intervals and construct a (normalized) frequency distribution $P(E_S)$
of the energy of the subsystem, see Fig.~\ref{distribution}.

The hypothesis that Newtonian
dynamics causes the subsystem to visit points in phase-space
with frequencies that match with those of the canonical probability distribution
can now be tested as follows.
According to statistical mechanics,
the distribution of energy in the canonical ensemble is given by~\cite{GREI97}
\begin{equation}
\label{cd1}
p(E)=g(E) e^{-E/T}/Z,
\end{equation}
where $T$, $E$, $g(E)$ and $Z$ are the temperature (in units of $k_B=1$), the total energy, the density of states and the partition function, respectively~\cite{GREI97}.
The function $p(E)$ has a maximum at some energy $E^\ast$, the most probable energy at the temperature $T$~\cite{GREI97}.
In the vicinity of $E^\ast$, we have~\cite{GREI97}
\begin{equation}
p(E)= A e^{a_2(E-E^\ast)^2+a_3(E-E^\ast)^3+a_4(E-E^\ast)^3+\ldots},
\label{cd2}
\end{equation}
where $A$ is a normalization constant and the coefficients $a_n=(1/n!)\partial^n S(E)/\partial E^n|_{E^\ast}$
are determined by the microcanonical entropy $S(E)$ of the subsystem~\cite{GREI97}.

For the two models considered in this paper, the coefficients $a_n$
are simple functions of $T$ and $N_S$ (see Supplementary Information).
Therefore, using $T$ and $E^\ast$ as adjustable parameters
a fit of Eq.~(\ref{cd2}) to the histogram $P(E_s)$
obtained from the dynamical evolution of the subsystem
yields an estimate of the temperature $T_S$ of the subsystem.
As shown in Fig.~\ref{distribution}, the simulation data for $P(E_S)$ (red lines)
and fitted $p(E)$ (black lines) are in excellent agreement, for both models alike.
In the thermodynamic limit ($N_E\rightarrow\infty$ before $N_S\rightarrow\infty$),
all but the quadratic term in the exponential can
be neglected and the distribution is Gaussian~\cite{GREI97}.
Therefore, for large $N_S$, $T_S$ and $E^\ast$ can be obtained
by fitting a Gaussian to $P(E_S)$ but from Fig.~\ref{distribution},
it is clear that for small subsystem sizes, $P(E_s)$ deviates
significantly from a Gaussian. However, taking into account the higher-order terms
in the expansion Eq.~(\ref{cd2}), the agreement between simulation data
and the prediction of statistical mechanics is excellent.
Repeating the simulations with different initial conditions (including different initial energies for
the subsystem or the environment) strongly suggests that this agreement is generic.

If the estimate $T_S$ is indeed the temperature of the subsystem,
the second central moment of $P(E_S)$ should be
related to the specific heat of the subsystem.
To check this, we define
\begin{equation}
\label{spheat}
C_S\equiv\frac{\langle E_S^2 \rangle-\langle E_S \rangle ^2}{T_S^2}
\end{equation}
where $\langle E_S \rangle$ and $\langle E_S^2 \rangle$ are the first
and second moment of $P(E_S)$, respectively.
Our numerical results (see Supplementary Information) are in excellent
agreement with canonical ensemble theory.

As a conclusive test, we perform microcanonical Monte Carlo simulations for both models (see Supplementary Information)
and obtain the distributions $P_{\mathrm{mc}}(E_s)$ of the subsystems.
The microcanonical Monte Carlo simulation generates statistically independent configurations of the whole system
strictly according to the microcanonical distribution but
samples the phase space in a completely different manner than does Newtonian dynamics.
The Kullback-Leibler distance $D(P_{\mathrm{mc}}(E_s);P(E_s))$
is a convenient measure to quantify the difference between the
two distributions $P_{\mathrm{mc}}(E_s)$ and $P(E_s)$~\cite{PRES03}.
In all cases, we find that the difference between $P_{\mathrm{mc}}(E_s)$ and $P(E_s)$
is very small (see Supplementary Information).
For instance, for the system of oscillators with $N_S=20$, $N=65536$ and $10^8$ samples,
we find that $D(P_{\mathrm{mc}}(E_s);P(E_s))\approx4\times10^{-2}$,
indicating that the probability that the two distributions are different is very small.

Having established that the interaction of the environment with
the subsystem causes both systems to equilibrate and also
drives the latter to its canonical state,
it becomes possible to derive from the Newtonian dynamics
alone, estimates for the equilibration time.
To this end we express the equilibration time
estimated from the simulations in physical units.
Typical frequencies of vibration in a solid are of the order of $10^{11}$Hz.
Using this number to set the scale of the frequency $\Omega$ in our model,
we find that equilibration takes of the order of $10^{-9}\;\mathrm{s}$.
Similarly, for the system of magnetic moments,
a realistic value of $J/k_B$ is of the order of $10\;\mathrm{K}$,
yielding an equilibration time of the order of $10^{-8}\;\mathrm{s}$.
Classical spin systems with Hamiltonians that encode frustration and/or disorder of regular
or random kind are however expected to exhibit larger, possibly much larger equilibration
time scales. The dynamical properties for subsystems of such theories under Newtonian evolution
are beyond the scope of the present work.

Even though the subsystems and the environments
which we have simulated are very small in the thermodynamic sense,
the subsystem and environment equilibrate on a nanosecond time scale.
Therefore, for an isolated nanoparticle of even a few thousand
atoms, an experimental probe that concentrates on only a few of those atoms
should yield data that follows the canonical distribution.

Calculations have been performed at JSC under project JJSC09.
MIK acknowledges financial support from FOM, the Netherlands.
MAN is supported in part by the National Science Foundation.
This work is partially supported by the Mitsubishi Foundation (SM) and NCF, the Netherlands (HDR).

\bibliography{../ref}

\end{document}